\documentclass{llncs}

\usepackage{natbib}
\usepackage{graphicx}
\usepackage{fancyhdr}
\usepackage{lastpage}
\usepackage{pdfpages}
\DeclareGraphicsExtensions{.pdf,.png,.jpg}
\usepackage{booktabs}
\usepackage{tabu}
\usepackage{longtable}
\usepackage{multirow} 
\usepackage{tabularx}
\usepackage{lscape}
\usepackage{enumerate}
\usepackage{tikz}
\usepackage{color}
\usepackage{url}
\usepackage{pbox}
\usepackage{todonotes}
\usepackage{pifont}
\usepackage{amsmath}
\usepackage{array}
\usepackage{datatool}
\usepackage{tablefootnote} 
\usepackage{units}
\usepackage{arydshln}

\begin{document}

\title{A short review and primer on online processing of multiple signal sources in human computer interaction applications}
\author{Jari Torniainen\inst{1} \and Andreas Henelius\inst{1}}
\institute{Quantified Employee unit, Finnish Institute of Occupational Health
\email{jari.torniainen@ttl.fi},\\
POBox 40, Helsinki, 00250, Finland}
\maketitle              

\begin{abstract}
The application of psychophysiological in human-computer interaction is a growing field with significant potential for future smart personalised systems. Working in this emerging field requires comprehension of an array of physiological signals and analysis techniques. 

Stream processing sytems (SPS) are emerging computational platforms that can be utilized in human-computer interaction for real-time analysis of high-volume multimodal signals.  Usage of complementary information contained in multiple signals is desireable as it can make HCI systems more robust. In this preprint we review existing software and hardware solutions for HCI-centric stream processing systems. The preprint also includes a brief introduction into SPS design considerations and structure from the perspective of analysing physiological signals.

This paper aims to serve as a primer for the novice, enabling rapid familiarisation with the latest core concepts. We put special emphasis on everyday human-computer interface applications to distinguish from the more common clinical or sports uses of psychophysiology.

This paper is an extract from a comprehensive review of the entire field of ambulatory psychophysiology, including 12 similar chapters, plus application guidelines and systematic review. Thus any citation should be made using the following reference:

{\parshape 1 2cm \dimexpr\linewidth-1cm\relax
B. Cowley, M. Filetti, K. Lukander, J. Torniainen, A. Henelius, L. Ahonen, O. Barral, I. Kosunen, T. Valtonen, M. Huotilainen, N. Ravaja, G. Jacucci. \textit{The Psychophysiology Primer: a guide to methods and a broad review with a focus on human-computer interaction}. Foundations and Trends in Human-Computer Interaction, vol. 9, no. 3-4, pp. 150--307, 2016.
\par}


\end{abstract}


\section{Introduction}
In the other sections of the primer \citet{cowley2016primer} we have covered a wide range of individual biosensors and areas of application for them as well as signal fusion. While we have thereby looked at it from a theoretical perspective, we have not yet addressed the topic of how to perform these fusion operations (feature extraction and classification) involving multiple signal sources in real time while leveraging modern stream processing methods. Therefore, the final subsection focuses on the implementation and technical aspects of real-time \textit{stream processing systems} for the online extraction and fusion of indices from streaming biosignals. The calculation of user indices from one or more signals is often realised by means of machine learning techniques such as a classifier. We address the integration of these indices into various applications also, by listing currently available hardware and software solutions. 

\section{An overview of stream processing systems}
Real-time data fusion requires a tool for combining multiple physiological signal streams and performing online extraction of features from raw signal data. Typical database management systems do not perform this task well, as the databases must be updated constantly for incoming data and the relevant operations tend to be slow. Also, long-term data storage is not required in most applications, since short fixed-length buffers cover the commonly used time ranges. Systems that take these factors into consideration are  generally referred to as \textit{stream processing systems} (SPSs). Their origins lie in the need for real-time processing of high-velocity and high-volume data. Example applications can be found in the field of data-mining: fraud detection, stock market analysis, and manufacture monitoring. Although dated, the most comprehensive survey of generic stream processing systems can be found in \citet{stephens1997survey}. Most existing stream processing systems are generic; i.e., they can be configured to process various kinds of data. The most prominent examples of this type of SPS are AURORA \citep{abadi2003aurora} and its continuation project BOREALIS \citep{abadi2005borealis}. The downside of a generic SPS is the overheads required for configuring the system for a specific task. For this reason, specialised, task-specific SPSs designed to deal with particular data types have been developed. Examples of task-specific SPSs used for physiological data include brain--computer interfaces and body area networks (see \citet{chen2011body}). 

\begin{figure}[!t]
   \centering
   \includegraphics[scale=0.50]{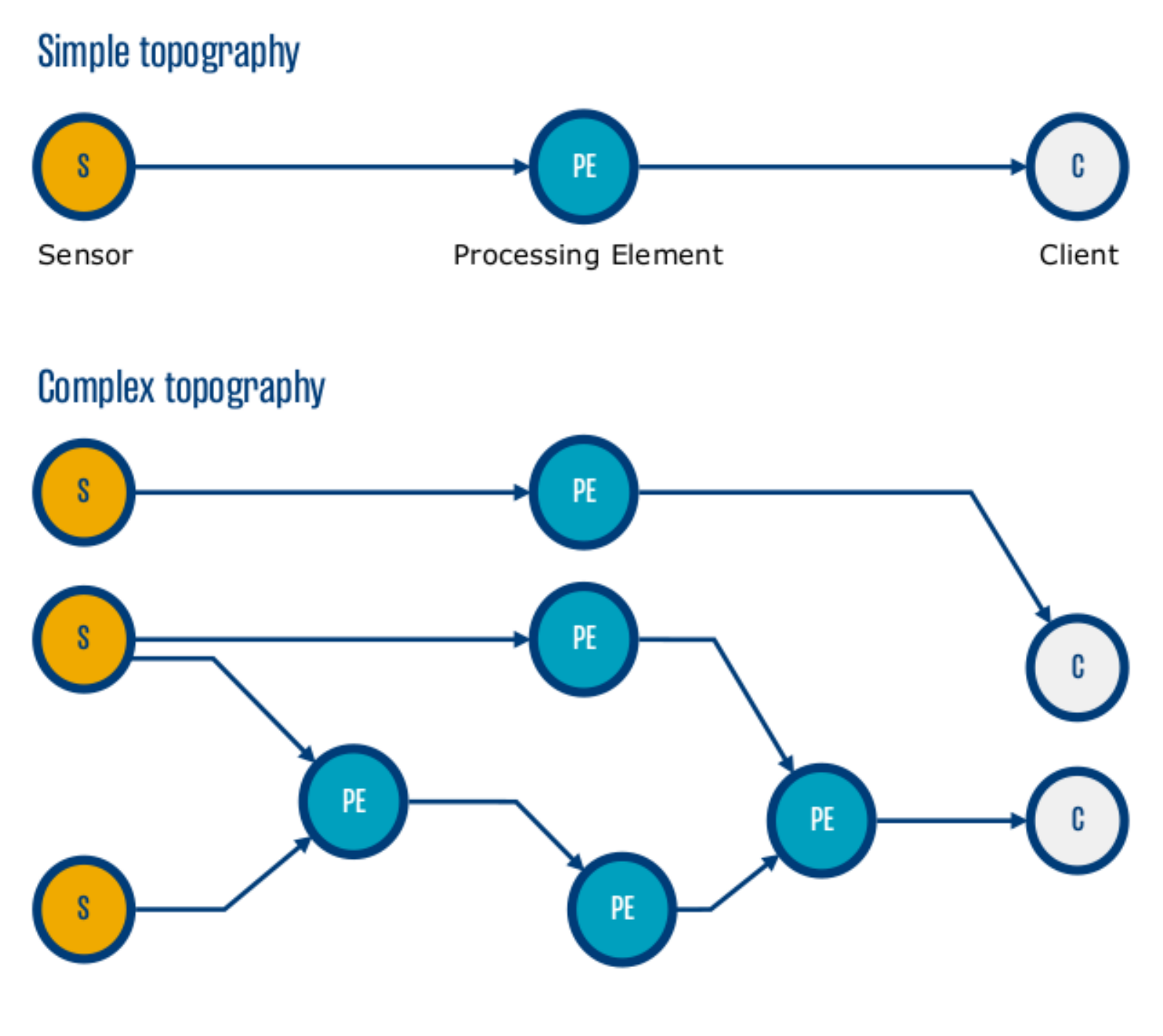}
   \caption{Two example topographies for stream processing systems.}
   \label{fig:sps}
\end{figure}

Stream processing systems are made up of three basic components, often termed \emph{sources}, \emph{filters}, and \emph{sinks} \citep{stephens1997survey}, although the naming conventions employed in the literature vary. For the sake of clarity, these components are referred to here as \emph{sensors}, \emph{processing elements}, and \emph{clients}, for consistency with the rest of the review. Stream processing systems are often represented as directed graphs comprising the three above-mentioned components. Two examples of SPS architectures, with different scales, are shown in Figure~\ref{fig:sps}. 

\subsection{Sensor}
Broadly speaking, one can define a sensor as an instrument used to measure some form of signal from a subject. In addition to the sensor element itself, this definition encompasses the interface and the protocol for transforming the incoming data into a format understood by the SPS. Sensor protocols include vendor-specific proprietary formats and APIs but also more open protocols such as the lab streaming layer\footnote{ See \url{https://github.com/sccn/labstreaminglayer}.} (LSL) \citep{kothe2013lab}. Most methods employed for recording physiology have some sensor instrument that can be implemented as part of an SPS.

\subsection{Processing elements}
Processing elements are the main computational units of an SPS. They can have multiple functions, but at the very minimum they perform some operation on incoming data. If the SPS is distributed in nature, the processing elements can be implemented on separate hardware, which provides for load balancing throughout the system. This is beneficial, as processing of signals with high sampling rates and high channel counts can be computationally intensive. Processing elements may also have internal buffers for the short-term storage of data. 

\subsection{Clients}
The term `client' describes any external application utilising information produced by the SPS. Clients can be considered end points (or sinks) in the SPS workflow. Some BCI frameworks consider the user of the system to be part of the client component. 

In addition to the basic components of an SPS, there are various properties that are intrinsic to most such systems. These include \emph{scalability}, \emph{distributed processing}, \emph{load management}, \emph{fault tolerance}, \emph{latency management}, and \emph{service discovery}. Scalability refers to the ability to extend the system to include more sensors and processing elements (scaling methods suitable for very large systems are discussed by \citet{jain2006design}). In distributed processing, different parts of the system can run on different hardware, possibly even in different (geographical) locations. Optimisation strategies for distributed SPSs are presented by \citet{golab2003issues}. The SPS must also apply some method of controlling multiple concurrent queries from multiple clients. The usual implementation is some form of load balancer (for details of various load management methods and fault tolerance, see \citet{abadi2003aurora, abadi2005borealis}). Since the system is composed of multiple, distributed parts, latency management is important; that is, it must control or monitor the latency between components. Various methods for achieving low-latency real-time operation exist in both the Aurora and the S4 \citep{neumeyer2010s4} system. Finally, to enable communication between different system components, automated discovery of various sources and processing elements should be part of the system. For instance, metadata-based identification and discovery of system components were used by \citet{aberer2007infrastructure}.

When one is fusing information from multiple signals for classification, it is important to consider the different time scales at which the signals operate too, as is noted by \citet{hogervorst:2014:a}. The robustness of index determination over time, as discussed by, for example, \citet{estepp:2011:a}, is an important issue, since the performance of the classifier degrades as time since calibration increases. Therefore, an online system should preferably incorporate some kind of automatic continuous calibration procedure.
Further design considerations, requirements, and guidelines for implementing SPSs are presented by \citet{balazinska2005fault, stonebraker2005eight, cherniack2003scalable}.

\section{Software solutions for online analysis of signals}
\label{sec:onlinesystems}
Several stream processing software solutions exist that are capable of gleaning knowledge about a user's cognitive state from biosignals. The various stream processing systems intended for psychophysiology are all similar in design, possessing the elements outlined in the previous section. There are, however, some differences with regard to what types of signals the tools primarily support and what constitutes their end-use purpose. We will briefly review some of the available stream processing systems intended for physiological data. The software covered in this section is presented in Table~\ref{sps_for_psychophys}

\begin{table}[ht]
\centering
\caption{Existing stream processing systems for physiological data}
\label{sps_for_psychophys}
\scriptsize{
\begin{tabular}{lll}
\textbf{Software} & \textbf{Reference} & \textbf{URL} \\
\midrule
BCI2000  & \cite{schalk2004bci2000}       & \url{http://www.schalklab.org/research/bci2000} \\
OpenViBE & \cite{renard2010openvibe}      & \url{http://openvibe.inria.fr/}                 \\
SSI      & \cite{wagner2011social}        & \url{http://hcm-lab.de/projects/ssi/}           \\
BCILAB   & \cite{kothe2013bcilab}         & \url{http://sccn.ucsd.edu/wiki/BCILAB}          \\
Wyrm     &                                & \url{https://github.com/bbci/wyrm}              \\
MIDAS    & \cite{midas}                   & \url{https://github.com/bwrc/midas}
\end{tabular}
}
\end{table}

Most of the SPS software packages available for physiological data are BCI and biofeedback frameworks. The BCI2000 package, for instance, is a full software solution tailored for BCI research. The design of BCI2000 is modular, cross-platform, and able to be extended with C++. OpenViBE is another cross-platform BCI solution. It is intended for biofeedback and neurofeedback and can be extended through C++ but also supports Python/Lua scripting. The SSI solution is a more versatile framework, suitable for processing a wide range of signals, from biosignals to video and audio. It includes functionality for feature extraction and classification. Extending SSI is handled with C++.  A BCI toolbox for MATLAB, BCILAB utilises the lab streaming layer protocol and provides a vast number of functions for analysing and classifying brain signals. Finally, the Python-based Wyrm is a more programming-oriented approach for implementing BCI set-ups. 

Recently, the MIDAS framework, developed for the online processing of signals, was introduced (by the authors of this section). The goal with MIDAS is to provide a cross-platform, generic framework that is easy to use and extend. While the MIDAS solution utilises LSL for signal input, it provides only the framework for constructing modular and distributed analysis systems. In other words, MIDAS contains only the building blocks for an SPS, and it is up to the user to implement the necessary analysis modules. The MIDAS framework is written entirely in Python and can be easily extended. Communication with clients takes place over a REST API. This makes it easy to integrate psychophysiological indices extracted online by means of MIDAS with other applications. 

It should be noted that BCI2000, OpenViBE, BCILAB, and SSI all can be set up and used without programming, since they include user-friendly tools for constructing workflows for BCI experiments. One key philosophy behind these software packages is to enable non-programmers to implement BCI/neurofeedback systems. In contrast, BCI systems implemented with the Wyrm toolbox or MIDAS require more programming and technical understanding. 

MIDAS and BCI2000 are the only stream processing systems reviewed here that support a distributed design. This attribute can be useful in processing tasks that involve multiple input signals and large volumes of data. 

\section{Hardware solutions for integrated online feature extraction}

In addition to software solutions for real-time signal fusion, there are dedicated hardware solutions for monitoring various aspects of a user's cognitive state; see Table~\ref{tab:hw_for_fusion}.

\begin{table}[ht]
\centering
\caption{Examples of hardware that performs online signal analysis}
\label{tab:hw_for_fusion}
\scriptsize{
\begin{tabular}{llll}
\textbf{Product and producer} & \textbf{Reference and/or URL} \\
\midrule
\textit{LifeShirt}                & \cite{coca2009physiological} \\  
& \url{http://vivonoetics.com/products/sensors/lifeshirt} \\
\textit{BioHarness 3}             & \url{http://www.zephyranywhere.com/products/bioharness-3} \\
Affectiva                         & \url{http://www.affectiva.com/}                            \\
Quasar                            & \url{http://www.quasarusa.com/technology_applications.htm} \\
B-Alert                           & \url{http://www.advancedbrainmonitoring.com/eeg-based-metrics} \\
JIN CO., \textit{JINS-MEME}       & \url{http://jins-meme.com/} \\ 
Optalert                          & \url{http://www.optalert.com/} \\
LumeWay                           & \url{http://www.lumeway.com/} \\
\textit{Bitalino-board}           & \cite{guerreiro2013bitalino}, \url{http://www.bitalino.com/} \\ 
\textit{e-Health Sensor Platform} & \url{https://www.cooking-hacks.com/}
\end{tabular}
}
\end{table}

For instance, the LifeShirt system is used to monitor the physiological signals (such as heart and respiration rate) of firefighters \citep{coca2009physiological}. Another example of hardware-based stream processing systems is the BioHarness 3, from Zephyr. Affectiva (founded by affective computing researcher Rosalind Picard) develops systems for online emotion recognition, especially using webcams (such as one called Affdex). Details of an interesting solution for monitoring heart rate (another application that uses a webcam) have recently been published \citep{poh:2010:a}.

Hardware solutions for monitoring cognitive states by means of EEG have been developed by Quasar and B-Alert. 
In addition, several solutions are available for tracking fatigue on the basis of eye tracking. Examples include JINS-MEME, Optalert, and LumeWay. 

Alongside application-specific products, there are platforms that allow a do-it-yourself approach to fusion of multiple biosensors. At the time of writing, the most noteworthy examples are the Bitalino-board \citep{guerreiro2013bitalino} and the e-Health Sensor Platform. Both feature numerous sensors (for EMG, EDA, BP, temperature, etc.) that can be connected to low-level microcontroller devices (e.g., an Arduino) or more high-level devices with compatible input/output connections (e.g., a Raspberry Pi). Both sets are highly configurable, with the trade-off that some expertise is required for building and operating systems on these platforms. 

\section{Conclusion}

The primary goal with online processing in HCI is to allow index determination to be used as a control signal for other components. Fusing information from multiple biosignals makes it possible to utilise complementary information and hence increase the robustness of the system. Achieving data fusion in real time is a challenging and computationally intensive task. However, that task can be addressed by utilising stream processing systems, such as those we have reviewed here.

\bibliographystyle{plainnat}
\bibliography{ch12_data_fusion_bib}
\end{document}